\documentclass[namedreferences]{SolarPhysics}
\usepackage{epsfig}
\usepackage{graphicx}

\newcommand{\be}{\begin{equation}}
\newcommand{\ee}{\end{equation}}
\newcommand{\beq}{\begin{eqnarray}}
\newcommand{\eeq}{\end{eqnarray}}

\begin{document}
\begin{article}
\begin{opening}

\title{Width of Radio-Loud and Radio-Quiet CMEs}

\author{G. \surname{Michalek}$^{1}$,
        N.   \surname{Gopalswamy}$^{2}$,
        H. \surname{Xie}$^{3}$
       }

\runningauthor{MICHALEK ET AL.} \runningtitle{Width of Radio-Loud
and Radio-Quiet CMEs}

\institute{$^{1}$ Astronomical Observatory of Jagiellonian
University, Cracow, Poland
                  \email{michalek@oa.uj.edu.pl}\\
           $^{2}$ Solar System Exploration Division, NASA GSFC,
Greenbelt, Maryland\\
           $^{3}$ Center for Solar and Space Weather, Catholic
University of America\\}

\date{Received ; accepted }

\begin{abstract}
In the present paper we report on the difference in angular sizes
between radio-loud and radio-quiet CMEs. For this purpose we
compiled these two samples of events using Wind/WAVES and SOHO/LASCO
observations obtained during 1996-2005. It is shown that the
radio-loud CMEs are almost two times wider than the radio-quiet CMEs
(considering expanding parts of CMEs). Furthermore we show that the
radio-quiet CMEs have a narrow expanding bright part with a large
extended diffusive structure. These results were obtained by
measuring the CME widths in three different ways.

\end{abstract}
\keywords{Sun: solar activity, Sun: coronal mass ejections, Sun:
solar  radio emission}

\end{opening}

\section{Introduction}
The relation between coronal mass ejections (CMEs) and type II radio
bursts has been studied for a long time, but is not fully understood
(see Gopalswamy  (2006) for a recent review). Properties of the
driving CMEs and the ambient medium through which the CMEs drive
shocks show a large variability, which seems to contribute to the
difficulties faced in understanding them (Gopalswamy {\it et al.},
2001). One of the major issues has been the lack of the type II
radio emission in the metric (Sheeley {\it et al.}, 1984) and
decameter-hectometric (DH) wavelengths (Gopalswamy {\it et al.},
2001) even for CMEs with speeds exceeding 1000 km s$^{-1}$.
Recently, Gopalswamy {\it et al.} (2007) performed a systematic
investigation of fast and wide (FW) CMEs that clearly lacked the
metric and DH type II radio emission (``radio-quiet''CMEs) and
compared them with the ones (``radio-loud'' CMEs) producing
detectable radio type II. It was found that the radio-quiet CMEs can
be distinguished from the radio-loud CMEs in three aspects: (1)
speeds and widths, (2) a fraction of halo CMEs, and (3) solar source
location of the CMEs. The radio-quiet CMEs are generally slower and
narrower than the radio-loud ones. The fraction of halo CMEs is much
larger for the radio-loud CMEs, which is related to the fact that
the radio-quiet CMEs are narrower on the average. It is also known
that halo CMEs are also faster and wider on the average (Yashiro
\emph{et al.}, 2004; Michalek, Gopalswamy, and Yashiro, 2003;
Gopalswamy, 2004). When the source locations were examined,
Gopalswamy (2006) and Gopalswamy \emph{et al.} (2007) found that
more than half of the radio-quiet CMEs were back-sided, while only a
small fraction (25) of the radio-loud CMEs were back-sided. They
attributed this result to the possibility that only a small fraction
of the shock surface is visible to the observer, thereby reducing
the possibility of detecting significant radio emission.  A fast but
narrow CME may have a similar limitation because the CME
cross-section and hence the shock surface area are expected to be
smaller. One of the suggestions made in Gopalswamy \emph{et al.}
(2007) is that most of the radio-quiet CMEs may have a narrow bright
part with extended diffuse structure. The purpose of this paper is
to examine the evolution of the width of radio-quiet and radio-loud
CMEs and compare them to confirm the smaller width of CMEs as a
contributor to radio quietness.

In  Section~2 the procedure for obtaining  the samples of the
radio-loud and radio-quiet CMEs is presented. In this section three
different  methods for the determination of CME widths are also
explained. In Section~3, we use the measured CME widths to show the
spatial difference between the radio-loud and radio-quiet CME
populations.

\section{Data and Determination of CME Width}
We consider the width evolution of radio-loud and radio-quiet CMEs
in the period of time from the beginning of SOHO/LASCO observations
(1996) until the end of 2005.  For this purpose we used
observational data from two instruments: The Radio and Plasma Wave
Investigation on the WIND Spacecraft (Wind/WAVE, Bougeret \emph{et
al}., 1995) and the Large Angle Spectroscopic Coronagraph on the
SOHO Spacecraft (SOHO/LASCO, Brueckner \emph{et al.}, 1995). To get
more accurate estimation of spatial sizes of the radio-loud and
radio-quiet CMEs additional determination of their width were made.
Details of these measurements are described in the next two
subsections.
\subsection{Data}
 The radio
bursts were identified in the dynamic spectra of the Radio and
Plasma Waves (WAVES) Experiment. The radio-loud and radio-quiet CMEs
are the two subsets of fast and wide CMEs
(speed$\geq$~900km~s$^{-1}$ and width$\geq$60$^\circ$). Gopalswamy
\emph{et al.} (2007) used the speeds and widths available in the
on-line catalog (http://cdaw.gsfc.nasa.gov/CME$\_$list). Here we
measure the widths again especially to follow the evolution of the
width. There were 469 FW CMEs, of which 195 were radio-quiet and 274
were radio-loud. The full list of the radio-loud and radio-quiet
CMEs and their properties are presented in  Gopalswamy  (2007). The
list excludes the three-month period (Jun to October 1998) when SOHO
was temporary unavailable.

\subsection{Width determination} One of the  interesting facts is the
 size difference between the radio-loud and radio-quiet CMEs.
For this purpose we measured widths of CMEs  in three different
ways. In Figure~1, examples of width determination for typical
radio-loud (2002/05/22) and radio-quiet (2002/09/16) events are
displayed. First,   in the top two panels
 we explain  the method used  for measurements employed for SOHO/LASCO
catalog  (Yashiro {\it et al.}, 2004).  In this method the CME width
is measured in LASCO C2 field of view (FOV) when CMEs reach the
largest angular size. Such a situation normally takes place in the
late phase of the propagation of CMEs when their leading edges are
observed in  LASCO C3 FOV.  The middle two panels
  show  the second method of  width determination.
  The method is similar to the catalog method,  but now widths are
measured when the leading edges of CMEs reach the boundary  of LASCO
C2 FOV.  It is clear that during the expansion, especially for the
very fast events, coronal plasma is compressed. These disturbances
appear as bright structures in LASCO observations influencing the
width determination. These disturbances do not give any addition to
type II radio emission. To get a more accurate space dimensions of
the radio-loud and radio-quiet CMEs, we decide  to determine the
width only from the main body of CMEs. For this purpose we measured
the  width excluding  the disturbance part. Examples of such
measurements are presented in the  bottom two  panels in  Figure~1.
This is the third method.

\begin{figure*}
\vspace{18cm}\includegraphics{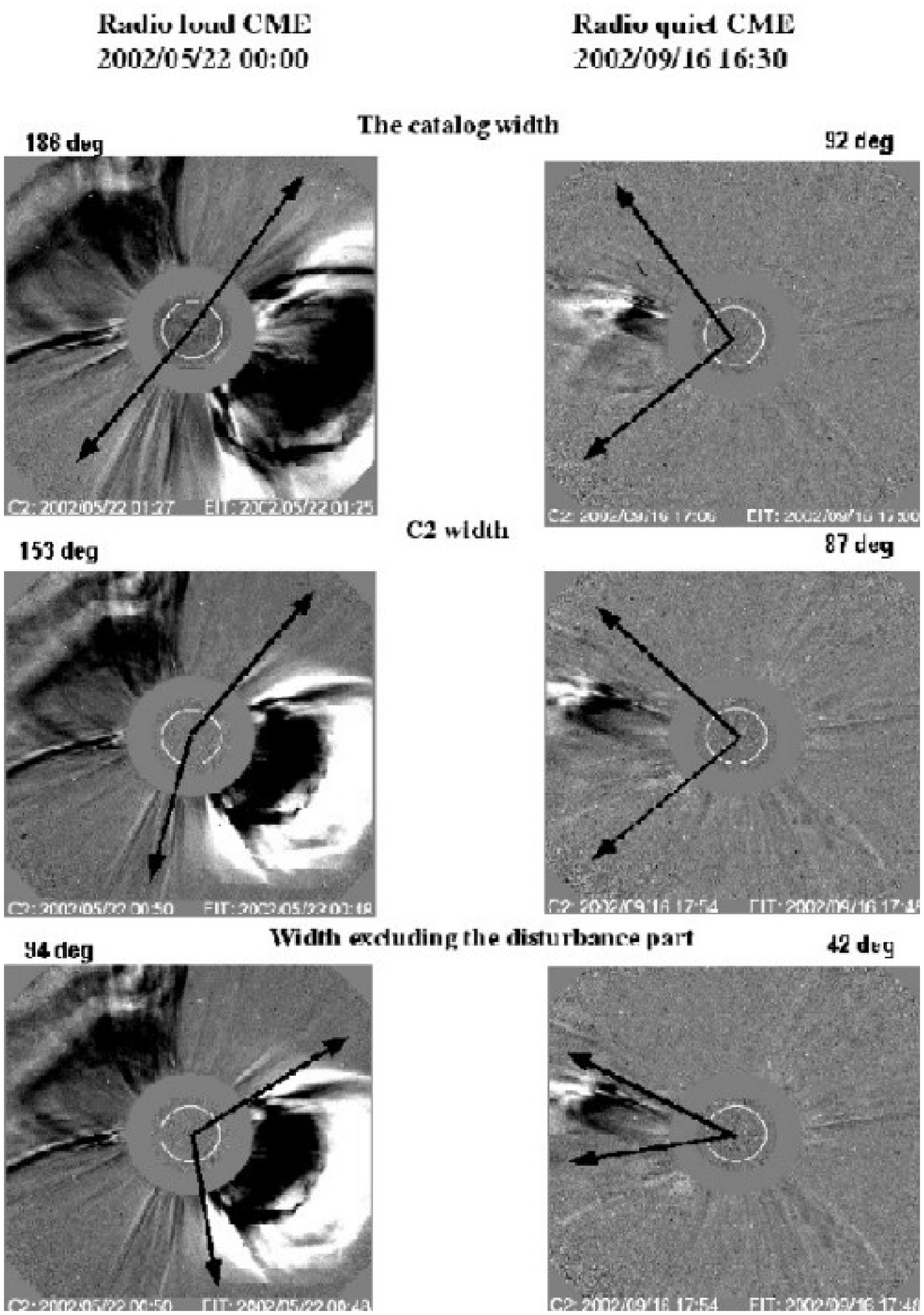} \caption{In the successive panels the three
methods of width determination for  examples of radio-loud and
radio-quiet CMEs are presented.}
\end{figure*}

\section{Results}
We compiled the  widths for all the  radio-loud and radio-quiet CMEs
using the three methods.  Figure~2  shows width distributions for
the radio-loud and radio-quiet CMEs taken from SOHO/LASCO catalog.
As we can see, in average, the radio-loud CMEs (the average width $=
138^{\circ}$) are about $10\%$ wider  than the radio-quiet CMEs (the
average width$=125^{\circ}$). To get more objective results we
compare widths excluding halo events.
In Figure~3, in two panels, width distributions measured in C2 field
of view are displayed. In this phase of evolution the radio-loud
CMEs (the average width$=116^{\circ}$) seem to be about $20\%$ wider
than the radio-quiet CMEs (the average width$=85^{\circ}$).
 For catalog and C2 measurements, widths
of CMEs are distributed over the whole possible angular range
($0^{\circ}-360^{\circ}$). Finally, we considered width
distributions  determined by excluding the disturbance part. The
distributions for the radio-loud and radio-quiet CMEs are shown in
Figure~4. As we may see, the radio-loud CMEs ( average
width$=69^{\circ}$) are  about $40\%$  wider than the radio-quiet
CMEs (average width$=47^{\circ}$). The expanding structures of the
radio-quiet CMEs are much narrower in comparison with the radio-loud
CMEs. Now, the measured widths are distributed over a smaller
angular range ($0^{\circ}-270^{\circ}$). It is interesting that the
expanding structures of CMEs are much narrower in comparison with
their total widths also. The ratios of the average catalog width to
the average  main body width are equal about 2.0 and 2.7 for
radio-loud and radio-quiet CMEs, respectively.

\setcounter{figure}{-1}

\begin{figure*}
\begin{subfigure}
\vspace{3.0cm} \includegraphics{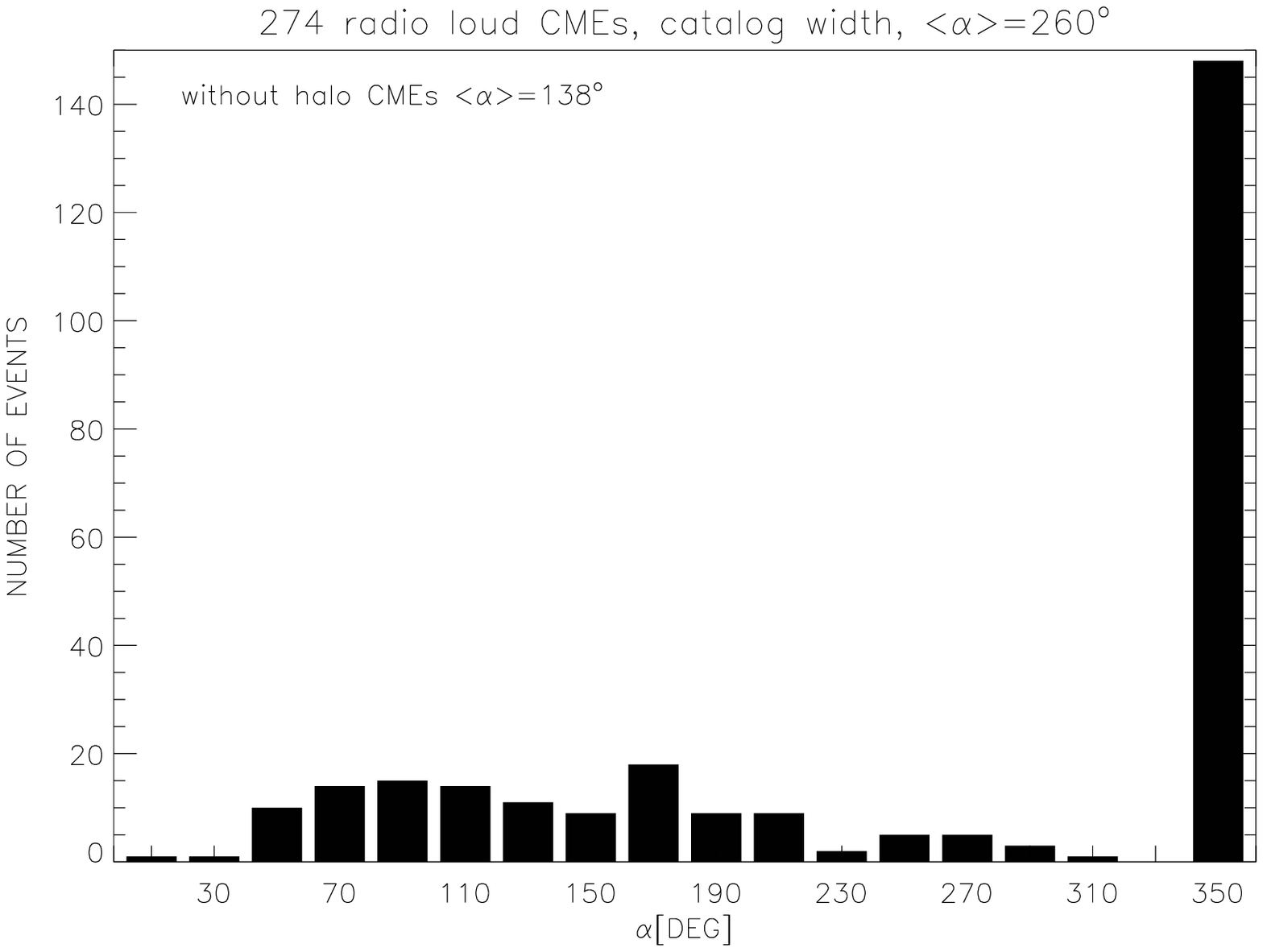} \vspace{3mm}


\end{subfigure}
\begin{subfigure}
 \vspace{3.0cm}  \includegraphics{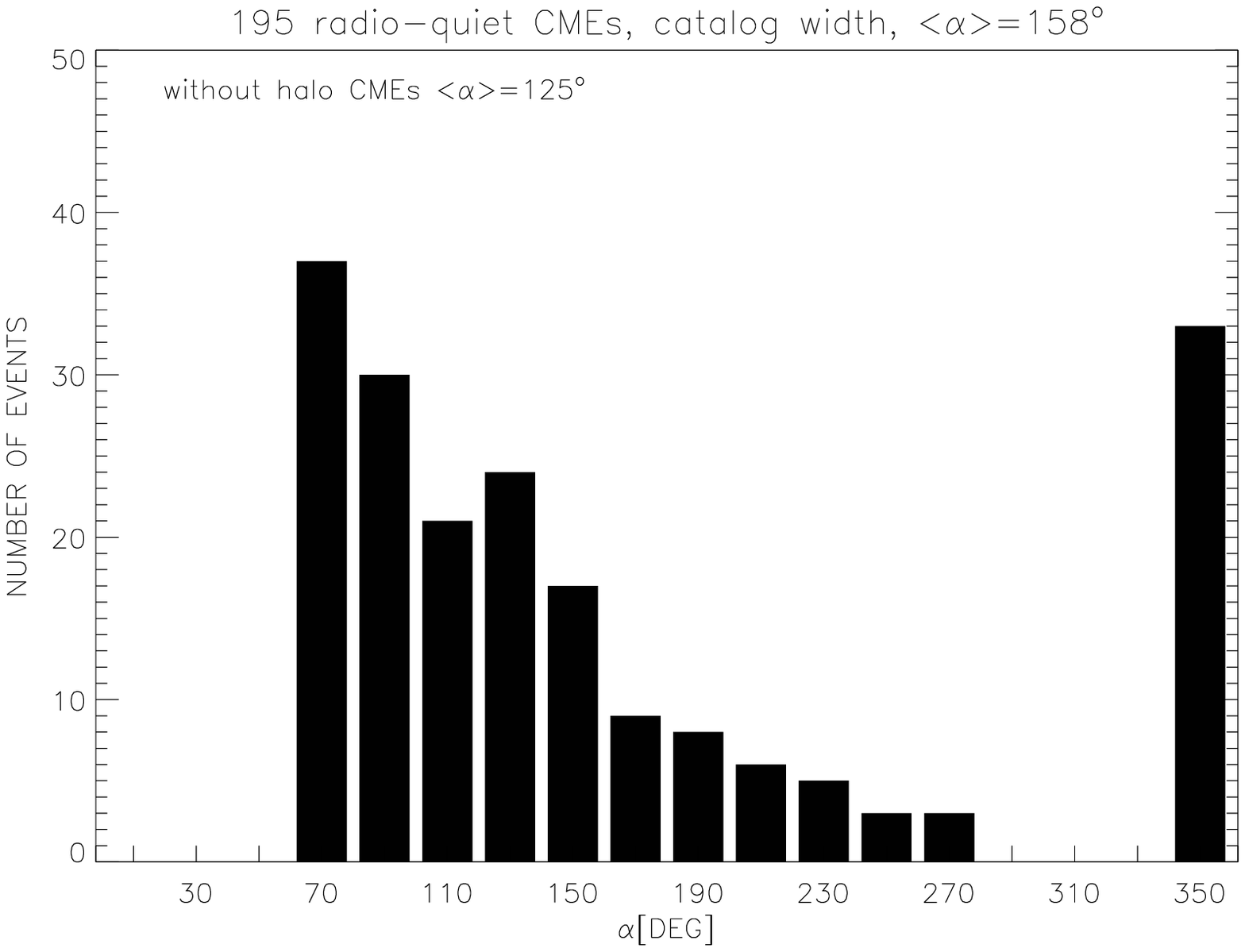}
\end{subfigure}
\caption{The distributions of CME width measured for SOHO/LASCO
catalog  for radio-loud (left) and radio-quiet (right) CMEs.}
\end{figure*}

\setcounter{figure}{0}

\begin{figure*}
\begin{subfigure}
\vspace{3.0cm} \includegraphics{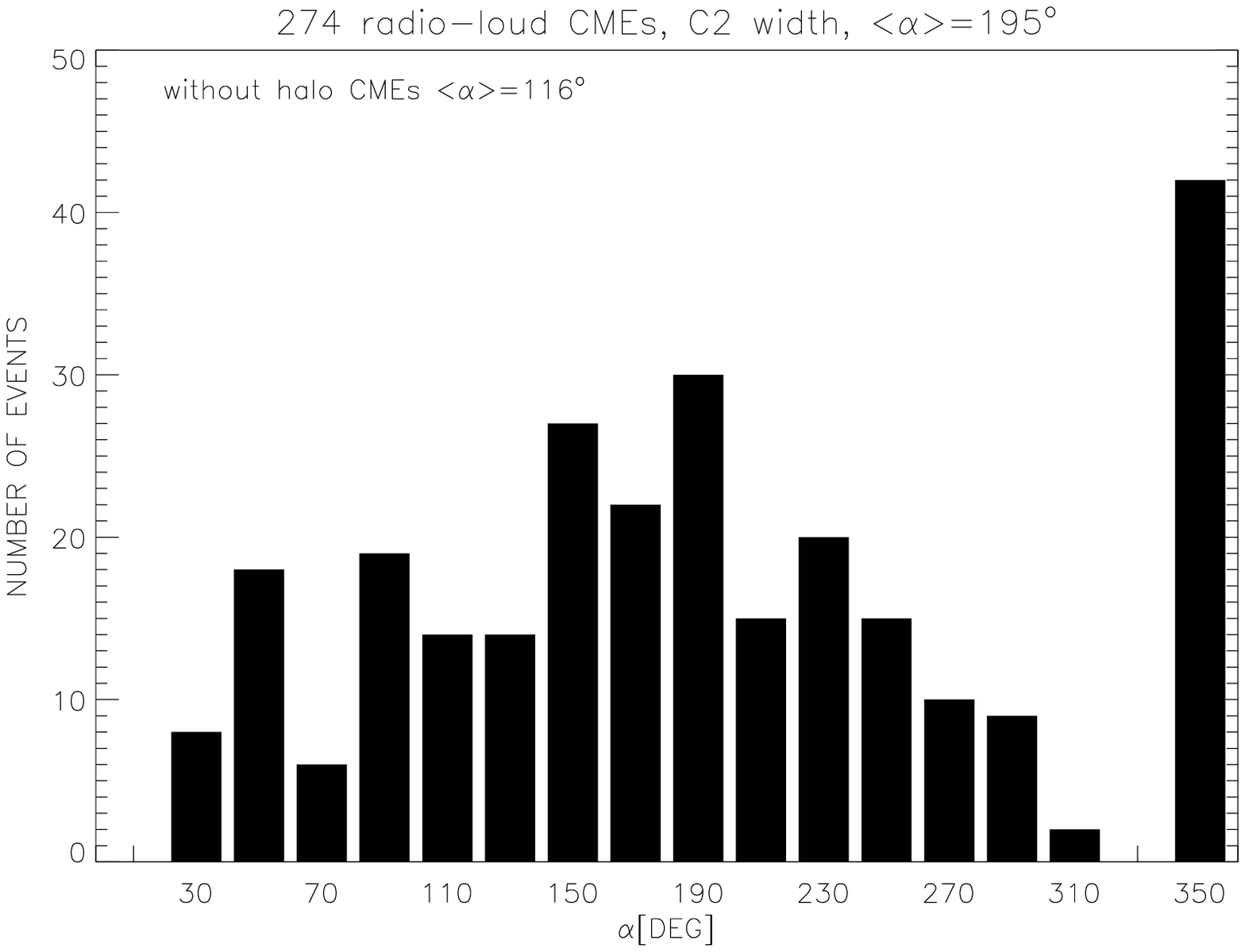} \vspace{3mm}


\end{subfigure}
\begin{subfigure}
 \vspace{3.0cm}  \includegraphics{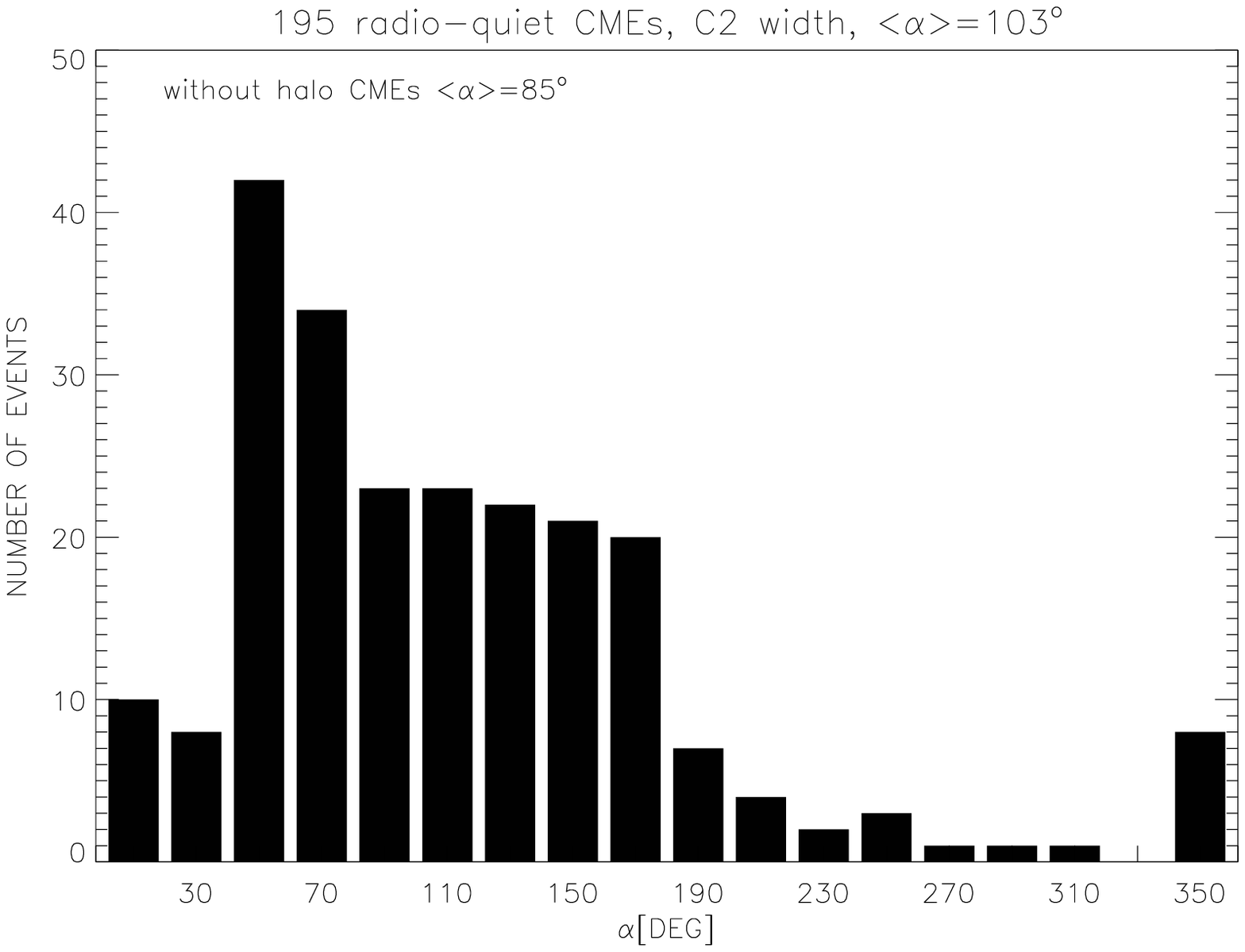}
\end{subfigure}
\caption{The  distributions of CME width measured during transit
from C2 to C3 LASCO fields of views for radio-loud (left) and
radio-quiet (right) CMEs. }
\end{figure*}

\setcounter{figure}{1}

\begin{figure*}
\begin{subfigure}
\vspace{3.0cm} \includegraphics{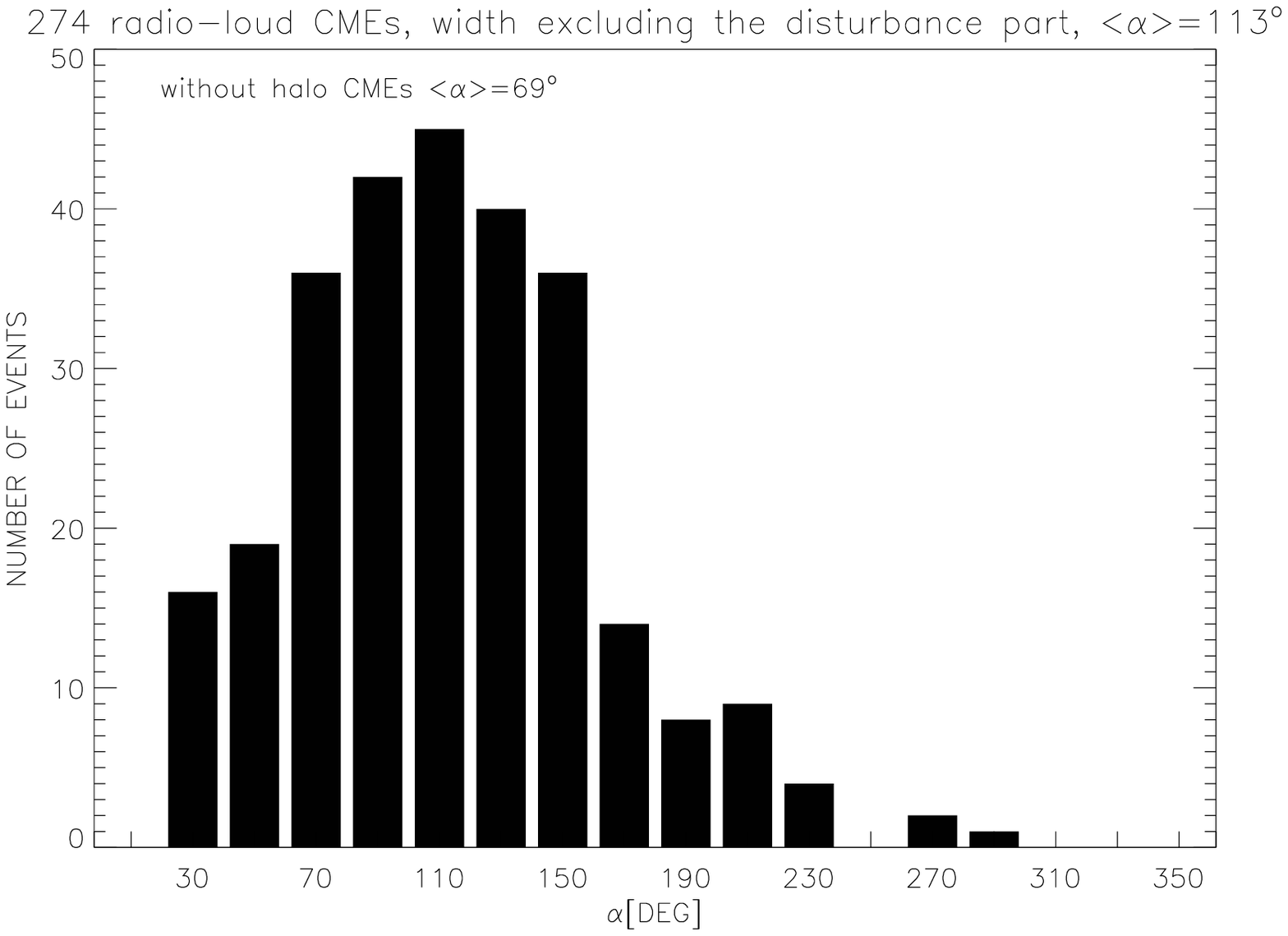} \vspace{3mm}


\end{subfigure}
\begin{subfigure}
 \vspace{3.0cm}  \includegraphics{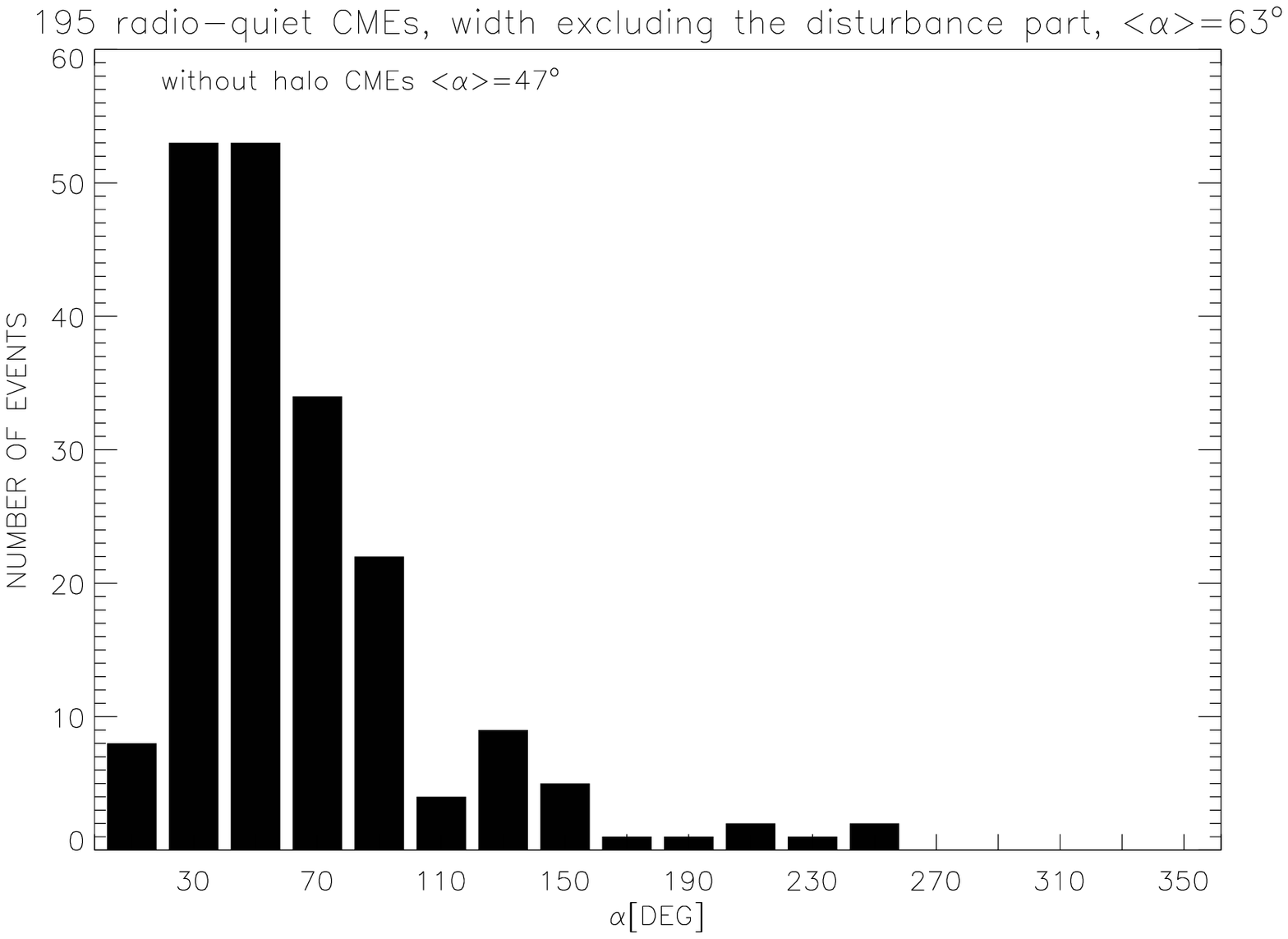}
\end{subfigure}
\caption{The  distributions of CME width excluding the disturbance
part for radio-loud (left) and radio-quiet (right) CMEs.}
\end{figure*}



\section{Summary}
In this study we examined the spatial difference between the
radio-loud and radio-quiet CMEs.  To get a reliable result, we
determined the angular widths of the radio-loud and radio-quiet CMEs
using three different methods. In all cases the radio-loud CMEs are
wider than the radio-quiet CMEs. When we compare the expanding parts
of CMEs (which are responsible for II type radio emission) the width
difference between the events is the largest. The expanding
structures of the radio-quiet CMEs are narrower ($\approx$40$\%$) in
comparison with those of the radio-loud CMEs.   The expanding
structures of CMEs are also much narrower in comparison with their
total widths, especially for the radio-quite events. The ratios of
the average catalog width to the average main body width are equal
about 2.0 and 2.7 for the radio-loud and radio-quiet CMEs,
respectively. The catalog widths for the radio-quiet CMEs are almost
three times bigger in comparison with the widths of expanding
structures. This means that the radio-quiet CMEs have a narrow
expanding bright part with a large extended diffusive structure. It
is clear that the spatial size of CME could be one of the most
important factors defining the presence of type II radio emission.
Our results proved the previous considerations ({\it e.g.}
Gopalswamy {\it et al.}, 2001, 2005; Pick and Maia, 2005;
Subramanian and Ebenezer, 2006). It is commonly accepted that type
II radio bursts are radio signatures of coronal MHD-shock waves
(Uchida, 1960; Wild, 1962). Flare-related blast waves and shock
driven by CMEs have been considered as two possible pistons of
metric type II bursts (see {\it e.g.} Cliver, Webb, and Howard,
1999), while the DH and longer wavelength bursts due to CME-driven
shock. CMEless type II bursts (Sheeley {\it et al.}, 1984) and the
discrepancy between the metric and IP type ( Reiner {\it et al.},
2001) were used to argue against the same shock causing the metric
and IP type bursts. Gopalswamy (2006) demonstrated that both these
discrepancies could be explained. It seems that CME-driven shock
works for the entire interplanetary space but additional mechanism
(blast waves) may operate for a narrow region ($\approx
1R_{\bigodot}$) close to the solar surface. In both cases, the width
of CMEs plays an important role in generation of fast particles and
radio bursts. Wider a given CME (more energetic event) wider a shock
front and larger area where particles can gain energy. Additionally,
larger CMEs could in bigger degree destruct  magnetic structures in
corona and amplify radio emission ({\it e.g.} Raymond {\it et al.},
2000; Pick {\it et al.}, 2006). This scenario is confirmed by strong
correlation between complex type III and type II radio bursts
associated with CMEs (Cane, Erickson, and Prestage, 2002;
Gopalswamy, 2004).
\begin{acknowledgements}
Work done by Grzegorz Michalek was partly supported by {\it MNiSW}
through the grant N203 023 31/3055.
\end{acknowledgements}
{}
\end{article}


\begin{thebibliography}{}
\bibitem [Bougeret et al., (1995)]{bou95}Bougeret, J.-L., Kaiser, M.L., Kellogg, P.J., Manning, R., Goetz, K., Monson, S.J.,
{\it et al.}: 1995, {\it Space Sci. Rev.}  \textbf{71}, 231.
\bibitem[Cane {\it et al.} (2002)]{can02}Cane, H.V., Erickosn, W.C.,
Prestage, N.P.: 2002, {\it J. Geophys. Res.}  \textbf{107(A10)},
SSH14-1.
\bibitem [Cliver {\it et al.}, (1999)]{cli99} Cliver, E.W., Webb, D.F.,
Howard, R.A.: 1999, {\it Solar Phys.} \textbf{187}, 89.
\bibitem [Brueckner {et al.}, (1995)]{bru95} Brueckner, G.E.,
Howard, R.A., Koomen, M.J., Korendyke, C.M., Michels, D.J., Moses,
J.D., \emph{et al.}: 1995 {\it Solar Phys.} \textbf{162}, 357.

\bibitem[Gopalswamy {\it et al.}, (2004)]{gop04} Gopalswamy, N.: 2004,
In:  Gary, D.E., and  Keller, C.U. (eds.), {\it Solar and Space
Weather Radiophysics: Current Status and Future Developments},
Kluwer Academic Publishers, Dordrecht, p. 305.
\bibitem [Gopalswamy {\it et al.}, (2006)]{gop06} Gopalswamy, N.: 2006,
In: Gopalswamy, N., Mewaldt, R., Torsti, J. (eds.), {\it Solar
Eruption and Energetic Particles}, {\it Geophys. Monograph}, Vol.
165, American Geophysical Union, p. 207.
\bibitem[Gopalswamy {\it et al.}, (2005)]{gop05} Gopalswamy, N.,
Aquilar-Rodriguez, E., Yashiro, S., Nunes, S., Kaiser, M., Howard,
R.A.: 2005, {\it J. Geophys. Res.} \textbf{110},  A12S07.
\bibitem[Gopalswamy {\it et al.}, (2001)]{gop01} Gopalswamy, N., Lara,
A., Kaiser, M.L., Bougeret,J.-L.: 2001, {\it J. Geophys. Res.}
\textbf{106}, 25261.
\bibitem[Gopalswamy {\it et al.}, (2007)]{gop01} Gopalswamy, N.,{\it et
al.}: 2007, {\it J. Geophys. Res.}, in press.



\bibitem [Michalek (2003)] {mic04} Michalek, G., Gopalswamy, N., Yashiro,
S.: 2003, {\it Astrophys. J.} \textbf{584}, 472.
\bibitem [Pick {\it et al.}, (2005)]{pik05} Pick, M., Maia, D.: 2005, {\it Adv. Space
Res.} \textbf{35}, 4430.
\bibitem [Pick {\it et al.}, (2006)]{pik06} Pick, M., Forbes, T.G., Mann, G.,
Cane, H.V., Chen, J., Ciaravella, A., {\it et al.}: 2006, {\it Space
Sci. Rev.} \textbf{123}, 341.
\bibitem [Raymond {\it et al.} (2000)]{ray00} Raymond, J.C., Thompson, B.J., St. Cyr, O.C., Gopalswamy, N.,
Kahler, S., Kaiser, M., {\it et al.}: 2000, {\it Geophys. Res.
Lett.} \textbf{27}, 1439.
\bibitem [Reiner {\it et al.} (2001)]{rei01} Reiner, M.J., Kaiser, M.L.,
Gopalswamy, N., Aurass, H., Mann, G., Vourlidas, A., Maksimovic, M.:
2001, {\it J. Geophys. Res.} \textbf{106}, 25279.

\bibitem[Sheeley {\it et al.}, (1984)]{she84} Sheeley, N.R., Howard,
R.A., Michels, D.J., Robinson, R.D., Koomen, M.J., Stewart, R.T.:
1984, {\it Astrophys. J}.  \textbf{279}, 839.
\bibitem[Subramanian {\it et al.}, (2006)]{sub06} Subramanian, K.R.,
Ebenzer, E.: 2006, {\it Astron. Astrophys.} \textbf{451}, 683.
\bibitem[Uchida (1960)]{uch60} Uchida, Y.: 1960, {\it Publ. Astron. Soc. Japan}
\textbf{12}, 376.
\bibitem[Wild (1962)]{wil62} Wild, J.P.: 1962, {\it J. Phys. Soc. Japan} \textbf{17},
249.
\bibitem[Yashiro {\it et al.}, (2004)]{yas04} Yashiro, S., Gopalswamy,
N., Michalek, G., St. Cyr, O.C., Plunkett, S.P., Rich, N.B., Howard,
R.A.: 2004,  {\it J. Geophys. Res.} \textbf{109}, A07105.


\end{thebibliography}
\end{document}